\documentclass[12pt,preprint]{aastex}

\slugcomment{Submitted to The Astrophysical Journal (Letters)}

\shorttitle{Proper Motions of the BN Object and the I Radio Source}

\begin{document}

\title{Proper Motions of the BN Object and the I Radio Source
in Orion: Where and When Did BN Become a Runaway Star?}

\author{Luis F. Rodr\'\i guez}
\affil{Centro de Radioastronom\'\i a y Astrof\'\i sica, UNAM,
Apdo. Postal 3-72, Morelia, Michoac\'an, 58089 M\'exico}
\email{l.rodriguez@astrosmo.unam.mx}

\author{Arcadio Poveda}
\affil{Instituto de Astronom\'\i a, UNAM,
Apdo. Postal 70-264, M\'exico, D. F., 04510 M\'exico}
\email{poveda@servidor.unam.mx}

\author{Susana Lizano}
\affil{Centro de Radioastronom\'\i a y Astrof\'\i sica, UNAM,
Apdo. Postal 3-72, Morelia, Michoac\'an, 58089 M\'exico}
\email{s.lizano@astrosmo.unam.mx}

\and

\author{Christine Allen}
\affil{Instituto de Astronom\'\i a, UNAM,
Apdo. Postal 70-264, M\'exico, D. F., 04510 M\'exico}
\email{allen@astroscu.unam.mx}

\begin{abstract}

We present absolute astrometry of the core of the Orion molecular cloud,
made with Very Large Array archive data taken over the last two decades.
Our analysis reveals that both the BN object and the radio source I
have proper motions: the BN object has a proper motion of
$12.6 \pm 0.6$ mas yr$^{-1}$  (corresponding to a velocity
of $27 \pm 1$ km s$^{-1}$ at an adopted
distance of 450 pc) to the northwest, while the radio source I
has a proper motion of
$5.6 \pm 0.7$ mas yr$^{-1}$  (corresponding to
a velocity of
$12 \pm 2$ km s$^{-1}$) to the southeast. The 
motion of the two sources is nearly antiparallel,
diverging from a point in between them, where they were located about
500 years ago. These results suggest that the BN object and the radio source I 
were part of a multiple young stellar system that disintegrated
in the recent past. 

\end{abstract}

\keywords{astrometry --- ISM: individual (\objectname{Orion}) ---
radio continuum: stars --- stars: pre-main sequence}

\section{Introduction}

The dynamical decay of triple and multiple young stellar systems, in which
one or more member stars are ejected from the group, is expected to
have profound effects in the star formation process (Reipurth 2000).
However, very few clear examples of fast-moving, runaway objects have been found
in association with regions of star formation. 

Perhaps the best example of a young runaway star is 
the Becklin-Neugebauer (BN) object in the Orion
region. This bright 2 $\mu$m source (Becklin \& Neugebauer 1967)
was first detected as a radio source by Moran et al. (1983).
From radio data taken from 1986.3 to 1995.0, Plambeck et al. (1995)
discovered that it exhibited a velocity in the plane of the sky of
$\sim 50~km~s^{-1}$ to the NW. In a recent study, Tan (2004)
proposed that the BN object was ejected from the Orion Trapezium
some 4000 yr ago. 
However, the possible existence of another runaway star (JW451; Jones \& Walker 1988) 
recently ejected from the
Orion Trapezium (Poveda et al. 2005) led us to revise the origin
of BN's large motions.
The BN object, the Kleinmann-Low nebula and 
their immediate environment are very suggestive of another trapezium similar 
to the visible Orion Trapezium, but still embedded in an obscuring envelope. In fact, the 20 $\mu$m 
brightness contours of the BN-KL area (Gezari, Backman, \& Werner 1998) show a morphology 
strongly reminiscent of that of the Orion Trapezium, particularly as it should have 
looked at 20 $\mu$m before the UV radiation from $\theta^1$ Orionis C had ionized and cleared the 
obscuring envelope. Also, the projected dimensions (d $\sim$ 15" ) and the bolometric 
luminosity of the BN-KL complex ($L \simeq 10^5~L_\odot$) are similar to those of the Orion 
Trapezium. Moreover, a number of compact radio and
infrared sources are detected in the BN-KL region (see Fig. 6 of Gezari et al. 1998
and Fig. 2 of Shuping, Morris, \& Bally  2004), 
indicating that there are many fainter stars deeply embedded there.
Furthermore, from a high resolution study of the region,
Shuping et al. (2004) find the star density in the region 
to be as high as, or even higher than, 
that of the Trapezium cluster. The strong similarities between the BN-KL complex 
and the Orion Trapezium motivated us to hypothesize that the former environment is also a 
propitious place to accelerate a runaway star like BN. In the
following we examine observationally the possibility that the BN object might have been accelerated 
by dynamical interactions occurring within a young multiple stellar system, 
still embedded in the obscured BN-KL region.
 
\section{Observations}

We searched in the archives of the
Very Large Array (VLA) of the National
Radio Astronomy Observatory (NRAO)\footnote{NRAO is a facility of the
National Science Foundation operated under cooperative agreement by
Associated Universities, Inc.} for observations of the Orion BN
region taken with high angular resolution
($0\rlap.{''}3$ or better), good coverage of the \sl (u,v) \rm
plane,  and that were made using as phase calibrator the nearby
($\simeq 1\rlap.^\circ6$) quasar 0541-056 (J2000 equinox).
We obtained data with these characteristics for five different
epochs: 1985.05, 1995.56, 2000.86, 2000.87, and 2002.25 (see Table 1).
To obtain reliable absolute astrometry, we used for all epochs
the most recent refined position of 0541-056 for 2000.25
($\alpha(J2000) = 05^h~ 41^m~ 38\rlap.^{s}083; 
\delta(J2000) = -05^\circ~41'~49\rlap.{''}428$).
We found significant proper motions only in the BN object and
in the radio source I. The observations of the nearby bright and
compact sources A, D, G, and H (see Garay et al. 1987 and Felli et al.
1993 for the nomenclature) 
did not show significant proper motions in right ascension or declination
at a typical 3-$\sigma$ upper limit of 2-3 milliarcsecond per year (mas yr$^{-1}$).
Since our data coverage had a large gap between 1985.05 and 1995.56,
we used high angular resolution observations taken on 1991.67 and 1991.68
(see Table 1) using different phase calibrators (0501-019 and 0530+135 on
both epochs). Assuming that, on the average, the sources A, D, G, and H
did not have proper motions, we corrected systematically all positions
in the 1991.67 and 1991.68
images to make them consistent with the other epochs.
In summary, we analyzed seven epochs taken from 1985 to 2002, with
good sampling across this time period. 
The main observational parameters of these observations are
listed in Table 1. All the observations were made in the most extended
A configuration and those of 2000.86 and 2000.87 included the Pie Town link, that 
provides improved angular resolution. 

The continuum data were edited and calibrated following
the VLA standard procedures. The data for
1985.05, 1991.68, 1995.56, 2000.87 were self-calibrated, 
while those for 1991.67, 2000.86, and 2002.25, when the continuum was
in all cases observed simultaneously with bright maser emission, were
cross-calibrated, using the self-calibrated maser solutions
to calibrate the continuum. Cleaned maps were 
obtained using the task IMAGR of AIPS and the ROBUST parameter set to 0.

\section{Results}

Figure 1 shows the main result of our analysis. By comparing the first and
last epoch of the set of seven observations, we clearly see that
both the BN object and the radio source I have proper motions,
diverging from a point between them.
In Figure 2 we show the right ascension and declination of both
sources as a function of time.
The positions were determined from a
linearized least-squares fit to a Gaussian ellipsoid function using the task
IMFIT of AIPS.
The error in position for each observation
was taken to be the sum in quadrature of the relative error
of the source in the given image (proportional to the angular
resolution over the signal-to-noise 
ratio) and a systematic error, taken to be 10 mas 
in right ascension and declination at all epochs.

The absolute proper motions and the position angles of motion
are given in Table 2. We conclude that the BN object is moving
to the northwest with a velocity in the plane of the sky of
$27 \pm 1~km~s^{-1}$, while the radio source I is moving to
the southeast with a velocity in the plane of the sky of
$12 \pm 2~km~s^{-1}$. 
If the BN object and the radio source I were in the past part of a multiple stellar
system 
that became disintegrated some 500 years ago (see below), 
we expect that the ratio of velocities
in the plane of the sky, a factor of about 2.3, applies also to the radial velocities.
As discussed by Plambeck et al. (1995) and Tan (2004), the ambient molecular
cloud has an LSR velocity of about 9 km s$^{-1}$, while the velocity of the
BN object is $\simeq$21 km s$^{-1}$, giving a radial velocity
of $\simeq$+12 km s$^{-1}$ with respect to the
ambient cloud.
On the other hand, the mean velocity of the SiO maser associated with
the radio source I is $\simeq$5 km s$^{-1}$, giving a radial velocity
of $\simeq$-4 km s$^{-1}$ with respect to the 
ambient cloud. The ratio of radial velocities is $\sim$3 consistent with
the value of 2.3 determined from the velocities in the plane of
the sky. It should be pointed out, however,
that the SiO maser emission traces the interaction of a fast jet with
the surrounding medium and that the true radial velocity of the
source I may be difficult to recover from the velocity of the SiO
masers. We conclude that the BN object moves with a total velocity
of $\simeq$30 km s$^{-1}$, receding with respect to the
molecular cloud , while the radio
source I moves with a total velocity of $\simeq$13 km s$^{-1}$, approaching with
respect to the molecular cloud.  
The possibility that both the BN object and the radio source I had significant proper
motions had been mentioned by Plambeck et al. (1995), Tan (2004), and
Bally \& Zinnecker (2005), but this is the first time that direct
observational evidence is presented.

The relative astrometry between the BN object and the radio source I
is very accurate, since it does not include the estimate for the
systematic errors
that have to be taken into account for the absolute positioning of the sources.
We obtain
$\mu_\alpha = -12.1 \pm 0.5~ mas~yr^{-1}$ and
$\mu_\delta = 14.0 \pm 0.5~ mas~ yr^{-1}$, corresponding to
a total proper motion of $18.5 \pm 0.5~ mas~ yr^{-1}$
toward a position angle of $-40\rlap.^\circ8 \pm 1\rlap.^\circ5$, 
in excellent agreement with the determination of
Tan (2004) from independent millimeter observations
($18.1 \pm 2.2~ mas~yr^{-1}$ at a PA of $-37\rlap.^\circ7 \pm 5\rlap.^\circ0$).
Extrapolating the velocity vector
backward and assuming no acceleration, we find that the BN object and
the radio source I were within $0\rlap.{''}5$ of each other
(see Figure 3)
around epoch $1480 \pm 30$ (that is, about 500 years ago).

\section{Discussion}

We will consider two possibilities for the origin of the observed motions
of the BN object and source I. First, we will discuss the possibility
that they formed a triple system which interacted 500 years ago,
forming a very close binary (source I) and the runaway BN object.
To examine this possibility, we need 
to estimate the mass of both objects.
From mid-infrared imaging of the Orion BN/KL region
Gezari et al. (1998) found a luminosity for 
the bright core of the BN object of
2500 $L_\odot$ which corresponds to 
a central star with a mass
$M_{BN} \sim  8~  M_\odot$ (Schaller et al. 1992). 
On the other hand, if the ionized HII region around BN is
optically thin at $\nu \sim 216$ GHz (as proposed by
Plambeck et al. 1995) one
obtains an ionizing photon rate of
$\dot S \sim 4 \times 10^{45}~ {\rm s}^{-1}$ (Mezger, Smith, \& Churchwell
1974) that corresponds
to $M_{BN} \sim 12.6 ~ M_\odot$ (Diaz-Miller, Franco, \& Shore 1998).
For $8~  M_\odot < M_{BN} < ~ 12.6~ M_\odot$,
conservation of linear momentum of the original triple system implies 
that for the velocity ratio of 2.3 between the BN object and source I, 
the total mass of source I should be in the range 
$18 M_\odot < M_I <  29 M_\odot$. If each member of the binary system in I 
has half the mass, the total luminosity of this source would be 
in the range $L_I \sim 8,300 - 4 \times 10^4 ~ L_\odot$. 
If the primary dominates the mass, the maximum luminosity of 
source I could be as large as $L_I \sim 10^5 L_\odot$, comparable
with the total luminosity of the region. Nevertheless, this latter 
configuration is dynamically less favorable for ejection.

Source I is detected only in radio continuum emission,
and it is not detected even
at mid-IR wavelengths (e.g., Plambeck et al. 1995;
Greenhill et al 2004). Luminosities of the order of $10^4~ L_\odot$ 
are expected due to its association with SiO masers (Menten \& Reid 1995). 
If the ionized region is optically thin at $\nu > 100$ GHz 
(see Figure 2 of Beuther et al. 2004) one can estimate the
mm flux expected from the hypothetical binary system.
If the binary system has 2 components each with a mass of the order
of $9~ M_\odot$, the predicted optically-thin free-free flux
density at $230$ GHz is 
$ S_\nu \sim 32~ mJy$, consistent with a turnover frequency of 100 GHz. 
On the other hand, if the binary components have masses
larger that $14~ M_\odot$ (using the upper limit of $M_{BN}$ or
assuming that the primary dominates)
the predicted flux is $ > 1~ Jy$, i.e., at least
a factor of 30 greater than the flux expected from a 100 GHz turnover 
frequency. In the latter case,
the observed relatively weak continuum emission could be due to 
the presence of an ionized jet
as in the case of IRAS~20126+4101 (Hofner et al. 1999)  where 
very high density circumstellar gas ``quenches'' the normal HII region.
The extinction to this source is estimated to
be very large, $A_V \sim 60 - 1000$ (Gezari et al. 1998; Beuther et al. 2004).
Thus, even a very luminous source could be hidden by the obscuring
envelope of source I.
It is important to emphasize that the results of Beuther et al. (2004)
are also consistent with a source that becomes optically thin at
much higher frequencies and that the free-free emission
of source I can be quite large in the sub-millimeter.

As discussed above, 500 years ago, the BN object was within less
than 225 AU of source I,
consistent with both sources being members of a compact stellar group.
We now assume that the original triple system had a total negative energy given
by a fraction $\alpha$ of the final binding energy of the binary system in I, 
$-\alpha (G M_p M_s)/2a$, where $M_p$ and $M_s$ are the masses
of the primary and secondary stars, respectively, and $a$ is the
semi-major axis of the binary.
Energy conservation implies that  
$(a/ AU) = 47.5 (1-\alpha)f (1-f) (M_{BN} / 12.6~M_\odot)$,
where $f$ is the fraction of
the total mass of the primary. 
For $f=1/2$ and $\alpha=0.1$, 
the semi-major axis is in the range
$ 7~ AU < a < 11~ AU$ for the range for $M_{BN}$ discussed above,
resulting in a period $ 4~ {\rm yr} < P < 7~ {\rm yr}$.
The relative velocity of the stars at perihelion is
$v_p \sim \sqrt{G M_I / a}$, thus one expects 
$ 40~ {\rm km~ s}^{-1} < v_p < 70 ~ {\rm km~ s}^{-1}$.
The space velocities of the BN object and the I source indicate
that their trajectories make an angle of only $\sim 21^\circ$ with respect
to the plane of the sky.
The mass ratio of the binary 
components and
the angle of the orbital plane with respect to the sky are not
known. One could try to restrict these parameters looking for
evidence of velocity asymmetries or time variations 
of the observed distribution of 
velocity centroids of the SiO masers (Plambeck et al. 1990;
Menten \& Reid 1995).
However, as noted above, 
the SiO maser emission may not trace
the true radial velocity of the source I.
The main implication of the triple system scenario is that source I has to be
a massive binary.

A second possible scenario is that 
instead of having a triple system, the interaction ocurred between several members
of a collapsing protostellar group, as modeled by Poveda, Ruiz, \& Allen (1967).
In this scenario, 500 years ago at the position of ejection
(see location of black asterisk in Figure 3), a bound multiple protostellar group experienced 
close encounters which accelerated BN and I in opposite directions; in order to preserve 
momentum and the initial negative energy one or more of the protostars, 
possibly together with their disks and envelopes, recoiled as a tight binary and multiple 
moving objects
(see Fig. 2 in Poveda et al. 1967). This would explain why the region where BN and I 
were at minimum separation (see Figure 3), is at present 
nearly devoid of infrared sources. 
Note that at present the cluster of infrared sources IRc2 (ABCD)
is located within 1''-2'' to the northwest quadrant
of I, in the direction of the original position
of the proposed multiple cluster,
suggesting that the tight multiple 
system was initially composed of BN, I and the IRc2 complex, and that
the n-body interaction 
ejected I with a bound close companion together with the IRc2 objects,
thus explaining their relative positions at present. 
This multiple source
scenario implies that the members of the IRc2 group should be participating in
the expansion with a sizeable fraction of the velocity of I. This possibility
can be tested with future astrometric observations of the IRc2 cluster in the
infrared. This multiple source scenario would allow source
I to be less massive
than in the triple source scenario described before.

\acknowledgments

LFR and SL are grateful to 
CONACyT, M\'exico and DGAPA, UNAM for their support.
AP and CA are grateful to Alejandro 
Hern\'andez for his assistance.

\clearpage

\begin{figure}
\epsscale{1.1}
\plottwo{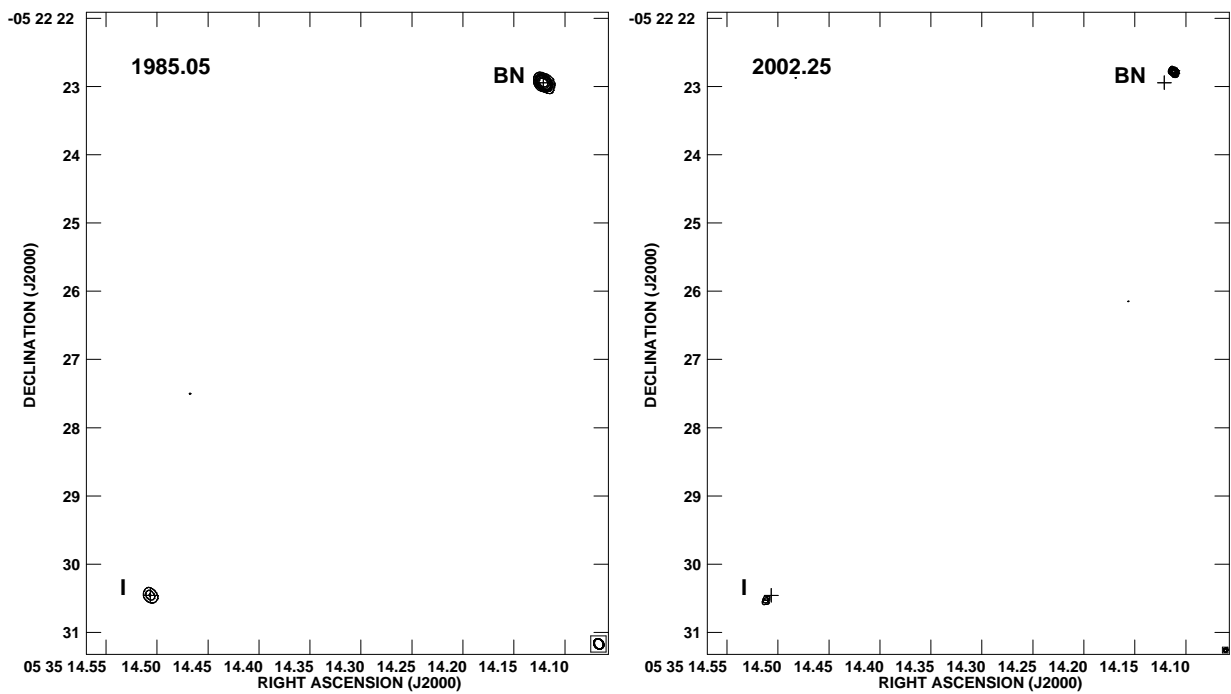}{f1b.eps}
\caption{VLA images for 1985.05 (left; 2 cm) and 2002.25 (right; 7 mm) towards 
the BN object and the I radio source.
The crosses mark the 1985.05 position of the sources.
Contours are -4, 4, 8, 12, 16, 24, and 32 times
0.13 (left) and 0.37 mJy beam$^{-1}$ (right), the rms of the respective
images. The half power contour of the synthesized beams
($0\rlap.{''}16 \times 0\rlap.{''}13$; PA = $32^\circ$ for the 
2 cm image and $0\rlap.{''}06 \times 0\rlap.{''}05$; PA = $-34^\circ$ 
for the 7 mm image) are shown in the bottom right corner of each panel.
Note the proper motion of both sources.
\label{fig1}}
\end{figure}

\clearpage

\begin{figure}
\epsscale{1.0}
\plottwo{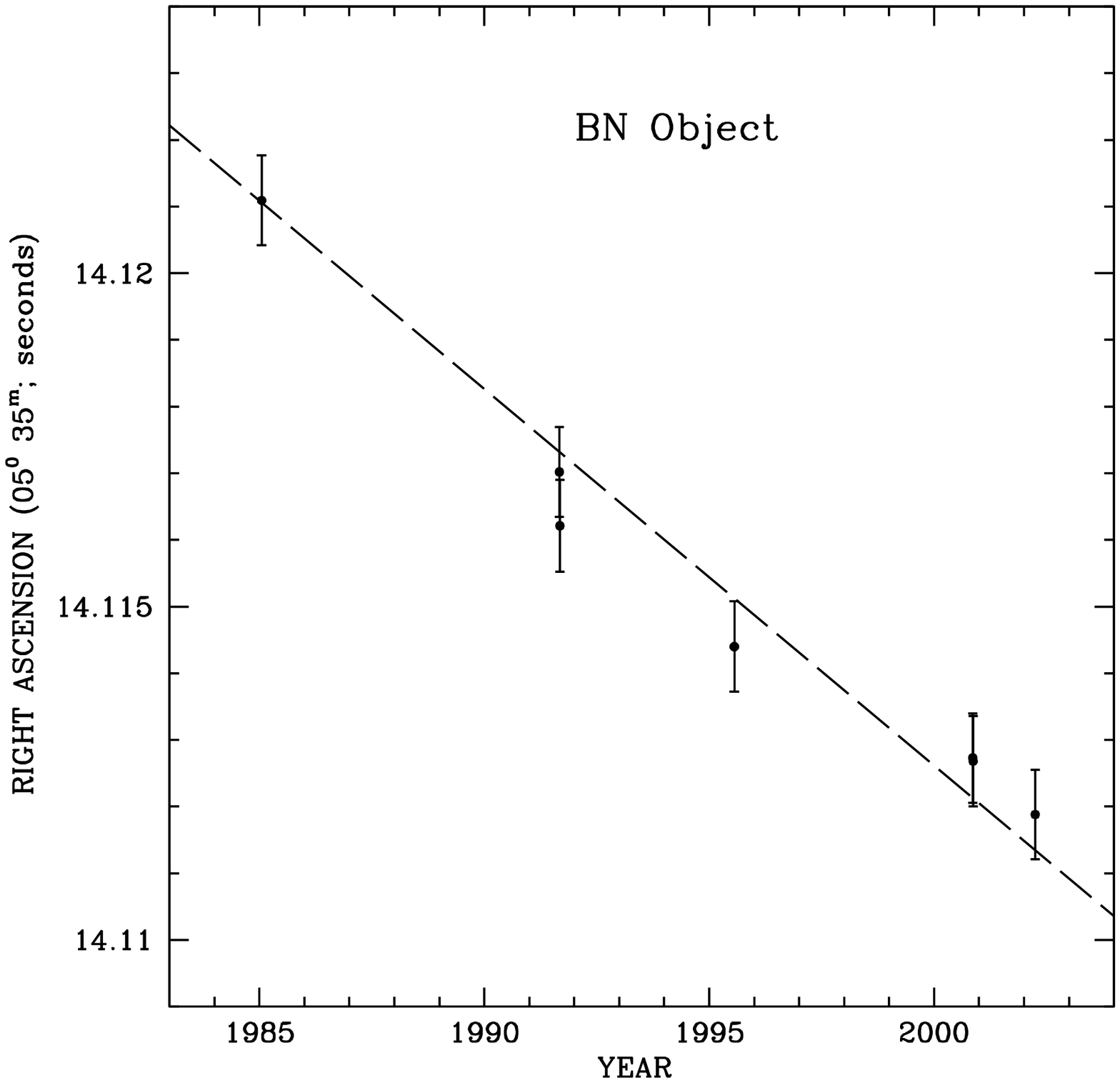}{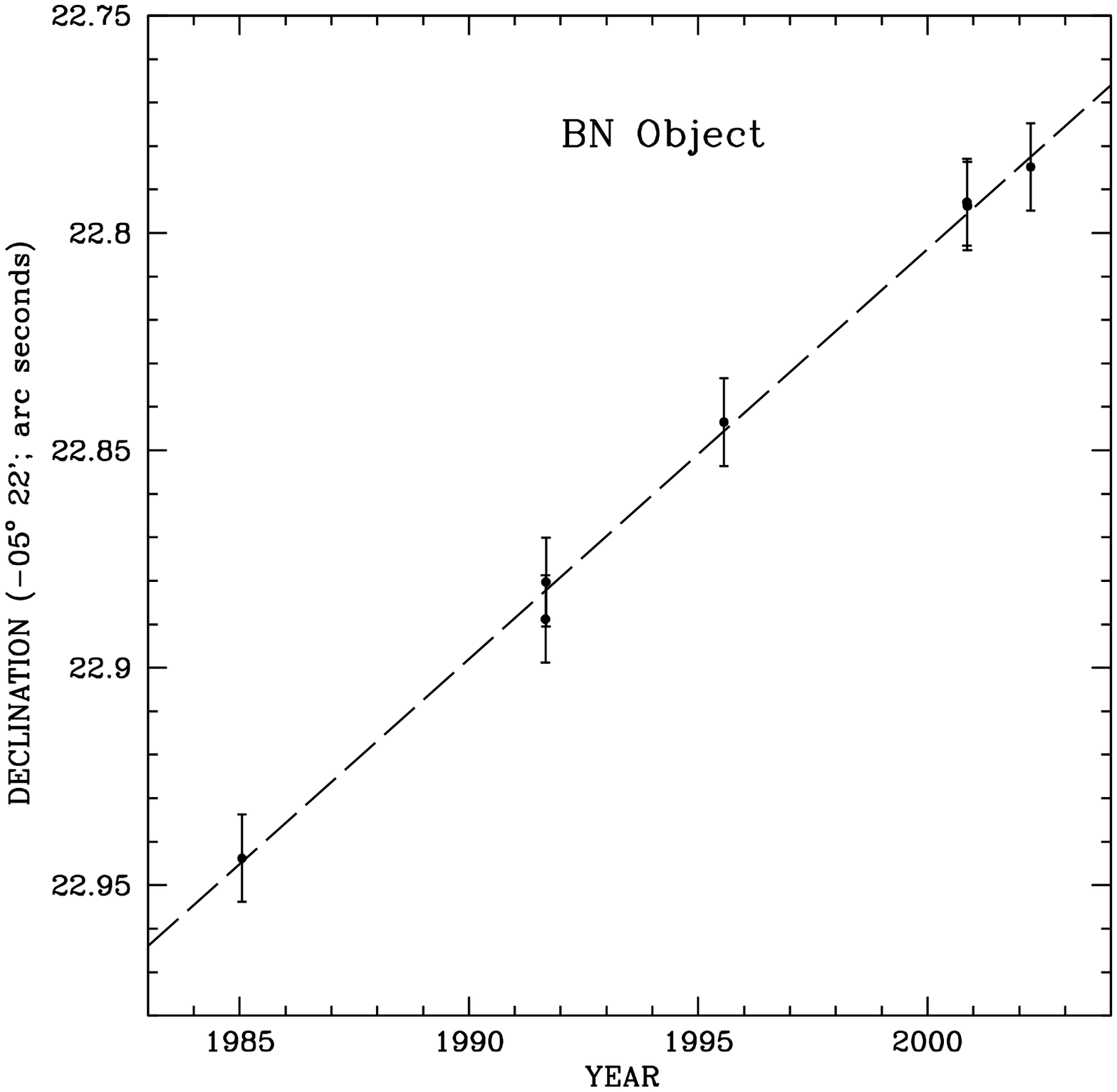}
\plottwo{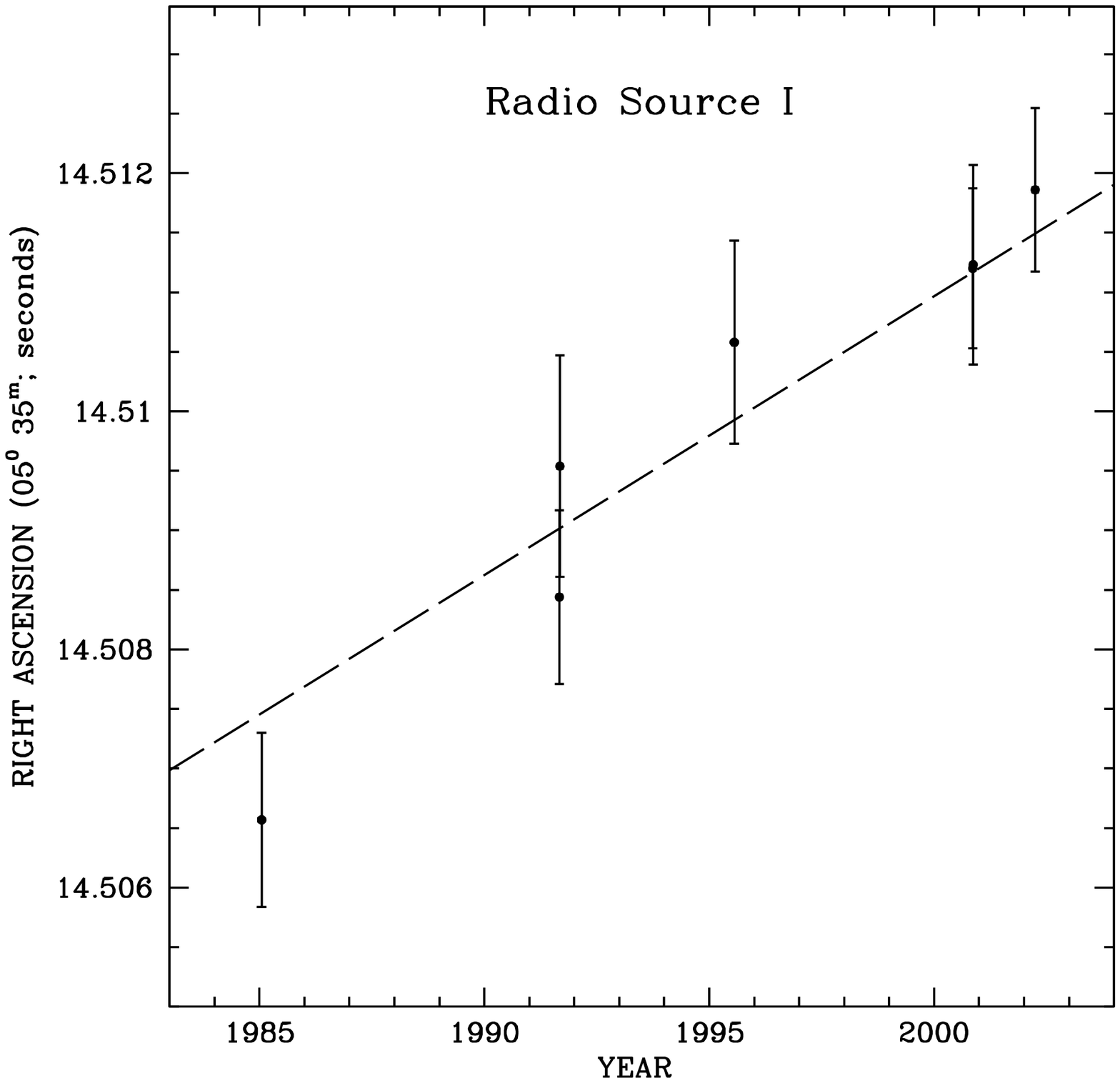}{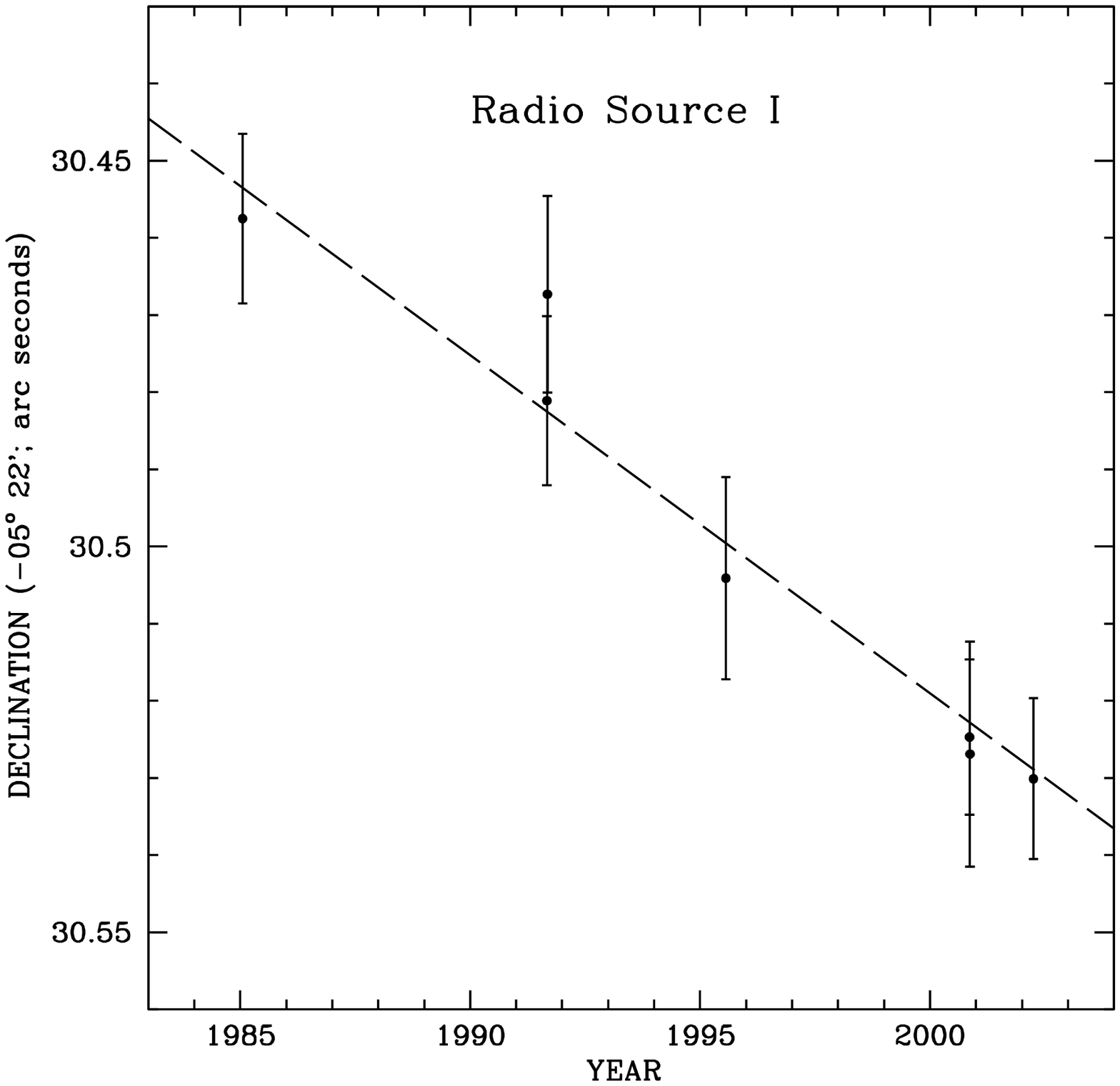}
\caption{Proper motions in right ascension (left) and declination
(right) for the BN object (top) and the radio source I (bottom).
The dashed line is the least squares fit to the data.
See Table 2 for the parameters of this fit.
The position of the BN object for epoch 2002.25 is
$\alpha(J2000) = 05^h 35^m 14\rlap.^s1119; \delta(J2000) = -05^\circ 22' 22\rlap.{''}785$,
while that of the radio source I is
$\alpha(J2000) = 05^h 35^m 14\rlap.^s5119; \delta(J2000) = -05^\circ 22' 30\rlap.{''}530$.
\label{fig2}}
\end{figure}

\clearpage

\begin{figure}
\epsscale{1.0}
\plotone{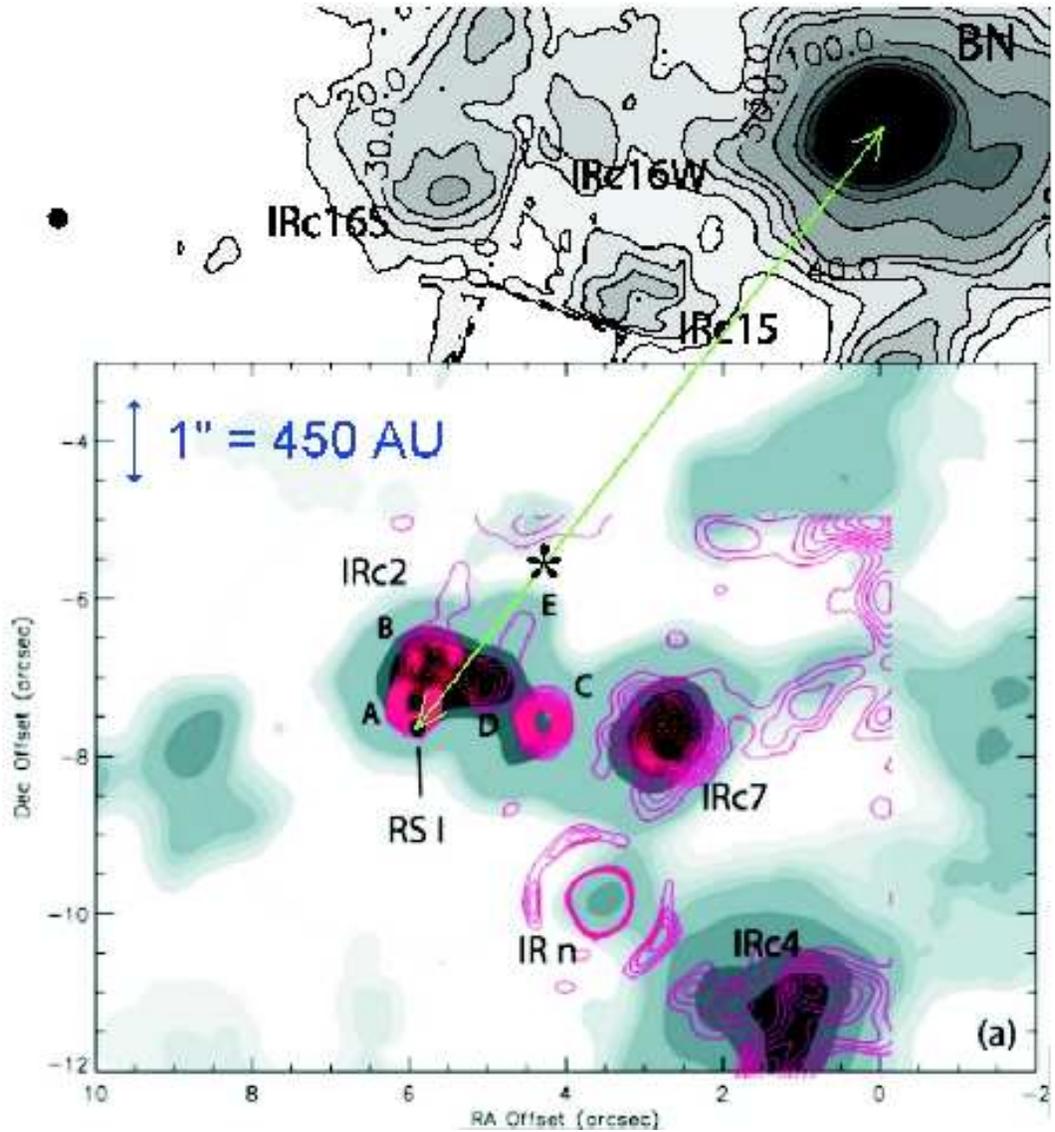}
\caption{Relative positions of sources BN, I, and IRC2 (ABCD).
The green line with arrows at its ends
connects the present positions of the BN object and the I radio source.
The black asterisk is the estimated position for the original location of
the multiple system from which BN and I were ejected to their present
velocities 500 years ago (see text).
Taken from Figs. 1 and 2a of Shuping et al. (2004).
\label{fig3}}
\end{figure}

\clearpage

\begin{deluxetable}{lccc}
\tablewidth{15.0cm}
\tablecaption{Archive Observations Analyzed}
\tablehead{
\colhead{}  & \colhead{Frequency} & \colhead{Wavelength} 
& \colhead{Synthesized Beam} \\  
\colhead{Epoch}  & \colhead{(GHz)} & \colhead{(cm)}
& \colhead{($\theta_M \times \theta_m; PA$)\tablenotemark{a}} \\
}
\startdata
1985 Jan 19 (1985.05) & 14.940 & 2.0 & $0\rlap.{''}16 \times 0\rlap.{''}13;~+8^\circ$ \\ 
1991 Sep 02 (1991.67) & 22.285 & 1.3 & $0\rlap.{''}10 \times 0\rlap.{''}10;~-16^\circ$ \\ 
1991 Sep 06 (1991.68) & 8.440 & 3.6 & $0\rlap.{''}26 \times 0\rlap.{''}25;~-55^\circ$ \\ 
1995 Jul 22 (1995.56) & 8.440 & 3.6 & $0\rlap.{''}26 \times 0\rlap.{''}22;~+34^\circ$ \\ 
2000 Nov 10 (2000.86) & 43.165 & 0.7 & $0\rlap.{''}04 \times 0\rlap.{''}04;~-14^\circ$ \\ 
2000 Nov 13 (2000.87) & 8.460 & 3.5 & $0\rlap.{''}24 \times 0\rlap.{''}14;~+27^\circ$ \\ 
2002 Mar 31 (2002.25) & 43.215 & 0.7 & $0\rlap.{''}06 \times 0\rlap.{''}05;~-34^\circ$ \\ 
\enddata
\tablenotetext{a}{Major axis$\times$minor axis in arcsec; PA in degrees}

\end{deluxetable}

\clearpage

\begin{deluxetable}{lccccc}
\tablewidth{15.0cm}
\tablecaption{Absolute Proper Motions}
\tablehead{
\colhead{}  & \colhead{$\mu_\alpha$} & \colhead{$\mu_\delta$}
& \colhead{$\mu_{total}$} & \colhead{Velocity\tablenotemark{a}} & \colhead{PA} \\
\colhead{Source}  & \colhead{($mas~yr^{-1}$)} & \colhead{($mas~yr^{-1}$)}
& \colhead{($mas~yr^{-1}$)} & 
\colhead{($km~s^{-1}$)} & \colhead{($^\circ$)} \\
}
\startdata
BN Object & $-8.4 \pm 0.6$  & $+9.4 \pm 0.6$ & $12.6 \pm 0.6$ & $27 \pm 1$ & 
$-42^\circ \pm 3^\circ$  \\
I Radio Source & $+3.5 \pm 0.7$ & $-4.4 \pm 0.7$ & $5.6 \pm 0.7$ & $12 \pm 2$ & 
$+141^\circ \pm 7^\circ$  \\
\enddata
\tablenotetext{a}{Velocity in the plane of the sky for an
assumed distance of 450 pc (Genzel \& Stutzki 1989).}

\end{deluxetable}

\clearpage


\end{document}